\documentclass{edp-conf}
\usepackage{graphicx}
\begin{document}

\TitreGlobal{SF2A 2002}

\title{Are X-ray flashes a peculiar class of (soft) gamma-ray bursts ?} 
\author{Daigne, F.}\address{CEA/DSM/DAPNIA, Service d'Astrophysique, 91191 Gif sur Yvette Cedex, France}
\author{Barraud, C.}\address{Observatoire Midi-Pyr\'{e}n\'{e}es, 14 Avenue Edouard Belin, 31400 Toulouse, France}
\author{Mochkovitch, R.}\address{Institut d'Astrophysique de Paris, 98 bis boulevard Arago, 75014 Paris, France}
\runningtitle{Are XRFs a peculiar class of (soft) GRBs ?}
\setcounter{page}{1}
\index{Daigne, F.}
\index{Barraud, C.}
\index{Mochkovitch, R.}

\maketitle
\begin{abstract} 
Heise et al. (2001) have reported the identification of a new class of sources by Beppo-SAX, the so-called X-ray flashes (XRFs), which have many common properties with $\gamma$-ray bursts (GRBs) but are not detected in the $\gamma$-ray range. 
In the framework of the internal shock model, we investigate the possibility that XRFs have the same origin than GRBs but are intrinsically softer due to 
different shock parameters.
\end{abstract}
\section{Introduction}
\vspace*{-1ex}

Heise et al. (2001) define XRFs with 4 criteria : (i) a strong non-thermal X-ray emission; (ii) a weak $\gamma$-ray emission; (iii) a duration less than a few 1000 s; (iv) no strong quiescent V/IR counterpart. 
In 5 years, Beppo-SAX detected 17 XRFs. Their distribution over the sky seems to be isotropic, their durations are comparable with those of GRBs detected by Beppo-SAX but their X- to $\gamma$-ray fluence ratios are higher. 
Since its launch, HETE-2 has also detected a few  XRFs. The preliminary study confirms that these events have many common properties with GRBs, except the fact that they are not detected above 50--100 keV.\\
\vspace*{-3.75ex}

\section{The internal shock (IS) model}
\vspace*{-1ex}

The IS model has been proposed by Rees \& M\'{e}z\'{a}ros (1994) and can reproduce the main temporal and spectral properties of GRBs (Daigne \& Mochkovitch, 1998). The $\gamma$-ray emission is due to relativistic electrons accelerated in shocks formed inside the relativistic outflow. The typical photon energy is
\begin{equation}
E_\mathrm{p} \simeq \left(1+z\right)^{-1} \times \Gamma \times E_\mathrm{p}'\ ,
\label{eq:ep}
\end{equation}
where $z$ is the redshift of the source, $\Gamma$ is the Lorentz factor of the emitting material and $E_\mathrm{p}'$ the typical photon energy in the comoving frame of the emitting material. As the radiative processes are still uncertain, we parametrize the energy $E_\mathrm{p}'$ by
$E_\mathrm{p}' \simeq K \times \rho^{x} \times \epsilon^{y}\ ,
\label{eq:epcom}$
where $\rho$ and $\epsilon$ are the density and the specific internal energy density in the shocked material and $x$, $y$ and $K$ are parameters depending on the radiative process.
The standard synchrotron radiation corresponds to $x=0.5$ and $y=2.5$ but smaller values of $x$ and $y$ are expected if, for instance, the fraction of accelerated electrons depends on the shock intensity, as suggested by Bykov \& M\'{e}sz\'{a}ros (1996). Daigne \& Mochkovitch (1998) indeed obtained synthetic lightcurves in much better agreement with the observations if $x=y=0.5$.\\
\vspace*{-3.75ex}

\section{Origin of the X-ray flashes}
\vspace*{-1ex}

Observations suggest a common origin for XRFs and GRBs. From eq.~\ref{eq:ep}, XRFs are either normal GRBs which are apparently soft due to a high $z$, or GRBs that are intrinsically softer (low $\Gamma\ E_\mathrm{p}'$).
The first possibility has to be rejected,
as XRFs and long GRBs have comparable durations.
 We test the second possibility using a simple version of the IS model. 
Two shells are ejected on a timescale $t_\mathrm{var}$,
with Lorentz factors $\Gamma_\mathrm{min}$, $\Gamma_\mathrm{max} = C\ \Gamma_\mathrm{min}$ and a constant injected power $\dot{E}$. Then,
\begin{equation}
E_\mathrm{p} \propto (1+z)^{-1} K \left(\dot{E} / t_\mathrm{var}\right)^{x}\Gamma_\mathrm{min}^{1-6x} f(C)\ ,
\label{eq:epis}
\end{equation}
where $f(C)$ is a steadily increasing function of $C$. 
Eq.~\ref{eq:epis} indicates that there are several ways of producing soft GRBs : (i) lower injected power $\dot{E}$; (ii) longer duration $t_\mathrm{var}$; (iii) less efficient shocks or less intense magnetic field (lower $K$); (iv) smoother outflow (lower contrast $C$);  (v) ``cleaner'' fireball, i.e. larger Lorentz factors (higher $\Gamma_\mathrm{min}$). \textit{Note that this last prediction of the IS model corresponds to GRBs with less baryonic pollution, which is just the opposite of a ``dirty'' fireball} (leading to a XRF in the external shock model).
Using the model developped in Daigne \& Mochkovitch (1998), we have carried out a set of simulations to test if (at least) one of these possibilities can explain the observed XRFs.\\
\vspace*{-3.75ex}

\section{Results and conclusion}
\vspace*{-1ex}

Our main results are : (i) a population of XRFs is indeed obtained if one adopt a broad range for the IS parameters : $\dot{E}$, $t_\mathrm{var}$, $K$, $\Gamma_\mathrm{min}$ and $C$; (ii) among all the possibilities listed above, \textit{small contrasts of the Lorentz factor} (low $C$) seem to produce the better agreement with the observed properties of XRFs. The conclusion is that \textit{in the framework of the IS model, XRFs are most likely soft GRBs associated with smooth outflows}. A possible observational test is that XRFs should have ``normal'' afterglows, but 
a low ratio of the afterglow over prompt energy outputs.
These results will be presented in details in a forthcoming paper.\\
\vspace*{-3.75ex}



\begin{thebibliography}{}
\bibitem{bykov:96} Bykov, A.M. and M\'{e}sz\'{a}ros, P., ApJ, 461, L37 (1996). 
\bibitem{daigne:98} Daigne, F. and Mochkovitch, R., MNRAS, 296, 275 (1998).
\bibitem{heise:01} Heise, J. et al., in GRBs in the Afterglow Era. Eds.  E. Costa et al. (2001).
\bibitem{rees:94} Rees, M.J. and M\'{e}z\'{a}ros, P., ApJ, 430, L93 (1994).
\end{thebibliography}
\end{document}